\documentclass[preprint,12pt]{elsarticle}

\usepackage{graphicx}
\DeclareGraphicsExtensions{.pdf,.png,.jpg}
\graphicspath{{fig/}{}}

\usepackage{color}
\usepackage{amsmath}
\usepackage{multirow}
\usepackage{booktabs}
\usepackage{algorithm, algorithmicx}
\usepackage{algpseudocode}

\usepackage{url}

\usepackage{amssymb}
\usepackage{listings}
\usepackage{tcolorbox}
\newtcolorbox{promptbox}{colback=gray!10, colframe=gray!50}
\journal{}

\begin{document}

\begin{frontmatter}



\title{Fault Cause Identification across Manufacturing Lines through Ontology-Guided and Process-Aware FMEA Graph Learning with LLMs}


\author[label1,label4]{Sho Okazaki\corref{cor1}} \ead{sho.okazaki.25@ucl.ac.uk}
\author[label2]{Kohei Kaminishi} 
\author[label1]{Takuma Fujiu} 
\author[label3]{Yusheng Wang}
\author[label4]{Manu Sasidharan} 
\author[label2]{Jun Ota} 

\affiliation[label1]{Department of Precision Engineering, School of Engineering, The University of Tokyo, 7-3-1, Hongo, Bunkyo-ku, 113-0033, Tokyo, Japan}
\affiliation[label2]{Research into Artifacts, Center for Engineering (RACE), School of Engineering, The University of Tokyo, 7-3-1, Hongo, Bunkyo-ku, 113-0033, Tokyo, Japan}
\affiliation[label3]{Institute of Engineering Innovation, School of Engineering, The University of Tokyo, 7-3-1, Hongo, Bunkyo-ku, 113-0033, Tokyo, Japan}
\affiliation[label4]{Department of Civil, Environmental and Geomatic Engineering, University College London, Gower Street, London WC1E 6BT, UK}

\cortext[cor1]{Corresponding author}

\begin{abstract}
Fault cause identification in complex engineered systems remains challenging due to system complexity, frequent reconfigurations, and the limited reusability of accumulated diagnostic knowledge, with automated manufacturing lines representing a prominent application domain. Although Failure Mode and Effects Analysis (FMEA) worksheets contain valuable expert insights, their reuse across heterogeneous system configurations is hindered by natural language variability, inconsistent terminology, and process differences. To address these limitations, we propose OGPAL (Ontology-Guided and Process-Aware Learning), a framework that enhances FMEA reusability by combining manufacturing-domain conceptualization with graph neural network reasoning. First, FMEA worksheets from multiple manufacturing lines are transformed into a unified knowledge graph through ontology-guided information extraction supported by a large language model (LLM), capturing domain concepts such as actions, states, components, and parameters. Second, a Relational Graph Convolutional Network (RGCN) with the process-aware scoring function learns embeddings that respect both semantic relationships and sequential process flows. Finally, link prediction is employed to retrieve and rank candidate fault causes consistent with the target line's process flow.

A case study on automotive pressure sensor assembly lines demonstrates that OGPAL outperforms a state-of-the-art retrieval-augmented generation baseline (nDCG@20 = 0.450) and an RGCN approach (0.559), achieving the best performance (0.719) in fault cause identification. Ablation studies confirm the contributions of both LLM-driven domain conceptualization and process-aware learning. These results indicate that the framework effectively supports reasoning over heterogeneous diagnostic knowledge and improves the transferability of FMEA knowledge across manufacturing lines, highlighting its potential applicability to a broader class of complex engineered systems.
\end{abstract}


\begin{highlights}
\item Reusing FMEA across heterogeneous engineered systems for fault cause identification.
\item LLM-guided information extraction enables lightweight domain conceptualization.
\item RGCN with a process-aware scoring function ensures consistent causal reasoning.
\item Best ranked cause retrieval: nDCG@20 of 0.719 vs. 0.450 and 0.559.
\item Enhances the reliability and adaptability of fault diagnosis across systems.
\end{highlights}


\end{frontmatter}

\section{Introduction}
\label{sec:intro}

Modern complex engineered systems, including automated manufacturing lines, consist of increasingly complex and diverse components \citep{xu2018industry}. While such systems enhance productivity and flexibility, their growing complexity makes fault diagnosis challenging and often requires sophisticated human expertise \citep{gonzalez2025enhancing}. Consequently, fault cause identification has become a fundamental challenge in complex engineered systems, with automated manufacturing lines representing a typical example \citep{aldrini2024fault}. For example, when assembly misalignment occurs in a manufacturing line, operators need to broadly consider multiple potential causes, such as incorrect robot teaching or loose fixed parts. However, non-expert operators often struggle to enumerate plausible fault causes due to limited experience and insufficient access to structured diagnostic knowledge \citep{vogl2019review}. Under such circumstances, it becomes essential to leverage expert knowledge accumulated in past fault analyses by integrating human experience with computational and knowledge-based technologies \citep{wan2023human,leo2019knowledge,burggraf2020knowledge}. By systematically drawing on such knowledge, diagnostic hypotheses can be generated more reliably, and candidate fault causes can be evaluated more effectively.

\begin{table}[!ht]
	\centering
\caption{FMEA Worksheet Example}
\label{tab:fmea_worksheet}
\begin{tabular}{lllll}
	\toprule
\begin{tabular}[c]{@{}l@{}}Process\\ description\end{tabular} & \begin{tabular}[c]{@{}l@{}}Failure\\ mode\end{tabular} & \begin{tabular}[c]{@{}l@{}}Failure\\ cause\end{tabular} & \begin{tabular}[c]{@{}l@{}}Failure\\ effect\end{tabular} & Recommendations \\ \midrule
\begin{tabular}[c]{@{}l@{}}Component\\ placement\end{tabular} & Misalignment & \begin{tabular}[c]{@{}l@{}}Incorrect\\ robot\\ teaching\end{tabular} & \begin{tabular}[c]{@{}l@{}}Assembly\\ failure\end{tabular} & \begin{tabular}[c]{@{}l@{}}Review robot\\ teaching\\ procedure\end{tabular} \\ \bottomrule
\end{tabular}
\end{table}

Among the various forms of expert knowledge, Failure Mode and Effects Analysis (FMEA) plays a vital role in systematic failure diagnosis. FMEA is typically performed during the design stage to anticipate potential failures and assess their causes and effects \citep{wu2021literature}. Beyond design, FMEA worksheets are also referenced during operation and maintenance \citep{case2010diagnostic}, and they provide structured records (see Table~\ref{tab:fmea_worksheet}) summarizing process descriptions (Function), possible failures (Failure), their underlying causes (Cause), and resulting effects (Effect) \citep{ebrahimipour2015ontology}. In conventional FMEA practice, failure modes are often associated with ordinal risk factors, such as Risk Priority Number (RPN), for failure prioritization. However, particularly during the initial development of FMEA for a manufacturing line, such scores are frequently preliminary or unavailable, and the analysis primarily relies on these qualitative descriptions. In this context, a ``line'' refers to a specific manufacturing line or system where FMEA is applied. A ``process'' describes the sequence of operations or tasks performed on that line, while a ``function'' represents the action of a specific process step. These steps are sequentially connected to form a process flow. For example, in an assembly line, process steps might include ``pick component,'' ``place component,'' and ``fasten component,'' each associated with a specific function. Failures are linked to these sequential functions, providing a structured way to analyze potential issues within the manufacturing workflow \citep{fujiu2026fbs}. While FMEA is usually prepared for individual manufacturing lines, the accumulated knowledge could serve as a transferable resource for diagnosing problems across heterogeneous lines.

However, direct reuse of FMEA knowledge across different manufacturing lines remains challenging, primarily due to various forms of heterogeneity inherent in manufacturing systems. First, manufacturing lines often exhibit structural heterogeneity, arising from differences in product configurations, equipment compositions, and functional decompositions, which leads to variations in how failures and their causes are represented. Second, semantic heterogeneity is prevalent across FMEA worksheets, where inconsistent terminology, line-specific failure naming, and implicit expert knowledge obscure shared fault patterns. Moreover, the tabular representation of FMEA further exacerbates this semantic heterogeneity, as relationships among cells are only implicitly encoded in rows and columns. Third, heterogeneity also extends to process order, as failures and their underlying causes are strongly conditioned by the sequence of operations. Even when similar failure phenomena are observed across different lines, variations in process order can result in distinct causal mechanisms. As a consequence, fault causes that should be comparable across lines become difficult to recognize, and reasoning based on isolated worksheets can lead to inconsistent or incomplete results. These forms of heterogeneity fundamentally constrain the transferability of FMEA knowledge and motivate the need for methods that explicitly address knowledge reuse under heterogeneous manufacturing contexts.

Previous studies have already alleviated the tabular limitations of FMEA. Ontology-based approaches have formalized worksheets into structured graph-based schemas \citep{mikos2011system,rehman2016ontology,camarillo2018knowledge}, and knowledge graph (KG) and machine learning (ML) approaches advanced reasoning capabilities. In particular, graph neural networks (GNNs) have been applied to learn relationships that reveal new links between elements which were not explicitly connected before \citep{filz2021data,luttermann2024automated}. Such relational learning is crucial when reasoning under heterogeneity. Retrieval-augmented generation \citep{lewis2020retrieval} has also been applied to retrieve fault causes based on natural-language similarity \citep{bahr2025knowledge}. Despite these advances, most research remains limited to single lines, and heterogeneity across products, equipment, terminology, and process flows continues to hinder knowledge transfer. There are studies that attempted to extend FMEA reuse across heterogeneous lines, including our previous works \citep{okazaki2023framework,fujiu2024description}, but these approaches rely on building heavy ontologies and use ad hoc reasoning, which reduces their practicality. These gaps indicate the need for a more generalizable approach. In particular, manufacturing knowledge should be conceptualized at the fundamental levels, such as actions, states, components, and parameters, that describe how processes operate and fail. Leveraging GNNs for consistent relational reasoning and incorporating process flow would then enable efficient FMEA knowledge reuse across heterogeneous manufacturing lines.

The purpose of this study is to enhance the cross-line reusability of FMEA knowledge to support fault cause identification in heterogeneous manufacturing lines. The key challenge is to generalize FMEA descriptions so that similar fault phenomena can be identified even when lines differ in equipment, materials, or terminology. To address this challenge, we propose OGPAL (Ontology-Guided and Process-Aware Learning), a framework that extracts domain-specific concepts such as actions, states, components, and parameters and align them through an ontology that provides a unified representation. In addition, differences in process flows between lines must be taken into account so that causal reasoning does not produce inconsistent results, although the modeling of process flow itself follows existing practice. Within OGPAL, GNNs are then employed to learn both the semantic relationships among ontology concepts and the constraints derived from process flow, enabling consistent reasoning across heterogeneous FMEA sources.

\section{Literature Review}
\label{sec:lit}

To address the low reusability of FMEA, prior research has proposed several approaches to reuse the knowledge described in worksheets. These approaches can broadly be divided into three categories: (1) ontology-based methods, (2) KG- and ML-based methods, and (3) RAG-based methods leveraging large language models (LLMs). Ontologies provide the schema and conceptual model, KGs represent instantiated knowledge with reasoning or learning, and RAG emphasizes semantic retrieval in natural language. Each category has expanded the scope of FMEA reuse, but important gaps remain.

\subsection{Ontology-Based Methods}
Ontologies are formally defined as an ``explicit specification of a conceptualization'' \citep{zuniga2001ontology}. In fault diagnosis, they have been widely used to describe core concepts and relationships \citep{liu2019ontology}, and to integrate heterogeneous data sources \citep{modoni2017enhancing}. In the context of FMEA, ontologies serve both as schemas and as graph databases for representing structured knowledge. Since FMEA worksheets are typically documented in natural language tables (see Table~\ref{tab:fmea_worksheet}), directly searching for relevant information is difficult. To overcome this, ontology-based studies have represented worksheet cells such as failure modes, causes, and effects as classes and relationships in a graph structure \citep{mikos2011system,rehman2016ontology,camarillo2018knowledge}. For example, an ontology schema may connect ``Failure\_mode'' to ``Mitigation\_action'' through the property ``hasMitigationAction,'' enabling structured queries and rule-based reasoning. Beyond manufacturing, ontology-based diagnostic frameworks have also been applied in other domains such as air conditioning systems \citep{hodkiewicz2021ontology} and wind turbines \citep{zhou2015research}. Several studies have extended ontologies with additional reasoning mechanisms, including Bayesian networks \citep{sayed2014ontology} and fuzzy theory \citep{ouyang2022multiple}. Although these works demonstrate the flexibility of ontologies as a foundation for hybrid diagnostic frameworks, they often remain limited to worksheet-level representations and do not explicitly model concepts such as actions, states, components, and parameters. As a result, their ability to generalize knowledge across different manufacturing lines with varying equipment and processes remains restricted.

To conceptualize failure phenomena in manufacturing systems, some studies have proposed expressive ontologies to capture domain-specific knowledge. In the design phase, systems are often conceptualized through functional, behavioural, and structural aspects, as formalised in the Function-Behaviour-Structure (Gero's FBS) framework \cite{gero1990design} and its extension \cite{goel2009structure}. Umeda et al. \cite{umeda1990function} also proposed the Function-Behaviour-State (Umeda's FBS) framework, which emphasizes the state aspect of systems and their transformations. Building on these representations, subsequent extensions explicitly link behaviours and functions to failure modes, enabling more detailed modelling of failure mechanisms and their propagation \cite{younus2025ontological,fujiu2026fbs}. In addition, several domain ontologies have been developed to represent manufacturing knowledge. Function System Ontology \cite{hodkiewicz2021ontology} connects FMEA concepts to upper-level ontologies and hierarchical system models, enabling reasoning about failure propagation across components. MALFunction Ontology (MALFO) \cite{compagno2025formal} defines failure-related concepts, such as malfunction, failure, and fault, in terms of their causal roles, providing a unified interpretation of heterogeneous failure descriptions.

While these frameworks provide strong theoretical grounding and expressive power, they are typically developed in a top-down manner and often require substantial manual effort in schema design, population, and alignment across systems. This poses practical challenges in heterogeneous manufacturing environments, where process structures and terminologies vary across manufacturing lines. In contrast, practical FMEA records exhibit recurring patterns in how failure phenomena are described, often combining action-oriented and state-oriented expressions linked to components and parameters. Our previous study \cite{okazaki2023framework} leveraged such patterns in a fault diagnosis system, demonstrating that such core information is  critical for representing operational conditions. These observations motivate a bottom-up abstraction of failure representations that focuses on fundamental concepts such as actions, states, components, and parameters, which can be more easily extracted from FMEA worksheets and aligned across heterogeneous lines.

\subsection{Knowledge Graph and Machine Learning-Based Methods}
Building on ontology schemas, more recent studies have instantiated FMEA knowledge as KGs and applied ML for automated reasoning \citep{wan2024making,simone2023industrial}. For example, Filz et al. \cite{filz2021data} reused FMEA knowledge with deep learning for maintenance optimization in aviation, Li et al. \cite{li2024combining} integrated FMEA and KGs for risk analysis of complex products, while Luttermann et al. \cite{luttermann2024automated} applied Markov decision processes to derive optimal treatments from FMEA in the medical domain. These works demonstrate the potential of combining FMEA with KGs to support decision-making across domains.

A critical step in these approaches is KG construction. In the context of FMEA, since conventional ontologies represent worksheet cells, prior studies have extracted individual cells as KG elements \citep{rehman2016ontology,camarillo2018knowledge}. For unstructured documents, natural language processing techniques are employed. Bidirectional Encoder Representations from Transformers (BERT) have been widely used for the extraction of cause-effect relationships in maintenance manuals \citep{liu2024task} and logs \citep{ding2025failure}. However, these methods usually require large amounts of annotated data that specify which words correspond to target concepts, which is both labor-intensive and inadequate for capturing manufacturing-specific terminology. More recent work leverages LLMs to expand existing concepts by prompting the model to generate related knowledge \citep{xia2024unlocking}. While these advances improve coverage, they still face challenges in generalizing knowledge across manufacturing lines that employ different concepts and terminology.

To utilize structured KGs, GNNs have been applied for reasoning tasks. GNNs have advantages over traditional ML methods because they can directly learn from graph structures and predict potential links between elements. Studies have used GNNs to infer potential cause-effect linkages in industrial systems \citep{mao2024graph}, predict maintenance dependencies among heterogeneous equipment \citep{xia2023maintenance}, and support decision-making with relation-aware architectures such as RGCN and RGAT \citep{zhou2021end,su2025knowledge}. Applications also include combining maintenance records with GNNs for recommendation \citep{mao2024graph} and modeling dependencies between sensor signals and fault types using GCNs \citep{liao2020fault}, and online disease diagnosis with KG-enhanced GNN reasoning \citep{wang2021online}. These studies highlight the increasing potential of graph learning for industrial fault diagnosis. However, existing GNN-based methods primarily capture local structural correlations, such as link prediction, and pay less attention to modeling process flows and sequential dependencies inherent in automated manufacturing lines. This limitation motivates the development of process-aware graph learning approaches that go beyond failure-related links to incorporate the sequential nature of manufacturing processes.

\subsection{LLM- and RAG-Based Methods}
The emergence of LLMs has further advanced FMEA reuse by enabling flexible handling of natural language. LLMs have been applied to structure general maintenance knowledge \citep{zhou2024causalkgpt,alvaro2025advanced} and to integrate with KGs for fault diagnosis \citep{peifeng2024joint}. RAG \citep{lewis2020retrieval} has been particularly promising. It retrieves documents relevant to a query and generates responses using an LLM. In industrial contexts, RAG has been explored for tasks such as digital twin support \citep{shi2025enhancing} and fault diagnosis \citep{nie2026industrial,gao2026knowledge,liu2025intelligent,miao2025graph}. For example, in CNC machining, LLMs have been used to extract structured information, such as symptoms, causes, and solutions, from unstructured maintenance records for diagnosis \citep{nie2026industrial,gao2026knowledge,liu2025intelligent}. Miao et al. \cite{miao2025graph} combined a KG constructed from maintenance logs with component-level information of bogies in high-speed trains, employing RAG for fault diagnosis. In the context of FMEA, Bahr et al. \cite{bahr2025knowledge} also combined FMEA KGs with RAG to retrieve semantically related fault causes, demonstrating higher reusability than direct worksheet search. These advances highlight RAG's strength in dealing with linguistic variability across fault descriptions. However, because retrieval is largely based on general language-level similarity, such methods cannot adequately capture the conceptual differences unique to manufacturing systems or the sequential dependencies among processes, which are critical for consistent causal reasoning in manufacturing.

\subsection{Research Gaps and Contributions}
While ontology-, KG- and ML-, and LLM- and RAG-based methods have advanced the reuse of FMEA knowledge, they remain insufficient for flexibly retrieving fault causes across heterogeneous manufacturing lines and for directly supporting operators. Existing methods (i) are tied to cell-level worksheet elements such as Failure, Cause, and Effect, which prevents generalization across lines that differ in equipment, terminology, and configuration, and (ii) often overlook process differences, which results in inconsistent reasoning across stages of production.

To address these limitations, the contributions of this study are as follows:
\begin{itemize}
    \item \textbf{Manufacturing-domain conceptualization.} We construct an ontology that shifts the representation of FMEA knowledge from worksheet elements to domain-level concepts including \textit{Action}, \textit{State}, \textit{Component}, and \textit{Parameter}. To make this step practical, LLMs are employed to automatically extract concepts from FMEA worksheets and instantiate them into a KG, avoiding the need for heavy manual ontology construction. This approach addresses limitation (i) and enables generalization of fault phenomena across heterogeneous lines.
    \item \textbf{Process-aware causal inference.} We incorporate information on process sequence into graph learning through the order-aware scoring function. To consider process dependencies, we define a distance metric based on the relative order of processes and integrate it into the link prediction objective. These mechanisms address limitation (ii), thereby preventing implausible predictions and improving the consistency of causal inference.
	\item \textbf{Practical framework for FMEA reuse.} We present an ontology-GNN framework that enhances the reusability of FMEA knowledge and demonstrates improved fault cause identification across multiple manufacturing lines, bridging the gap between academic proposals and industrial applications.
\end{itemize}

% Road map of the paper

The remainder of this paper is organized as follows. Section~\ref{sec:method} describes the proposed method, OGPAL, including the ontology design, LLM-driven KG construction, and process-aware graph learning. Section~\ref{sec:case} presents a case study on automotive pressure sensor assembly lines, demonstrating the effectiveness of the method compared to baselines. Section~\ref{sec:discussion} discusses the implications of the results and the limitations of the current approach. Finally, Section~\ref{sec:conc} concludes the paper by summarizing the key findings and contributions.

\section{Methodology: OGPAL}
\label{sec:method}

\begin{figure}[!ht]
		\centering
		\includegraphics[width=\textwidth]{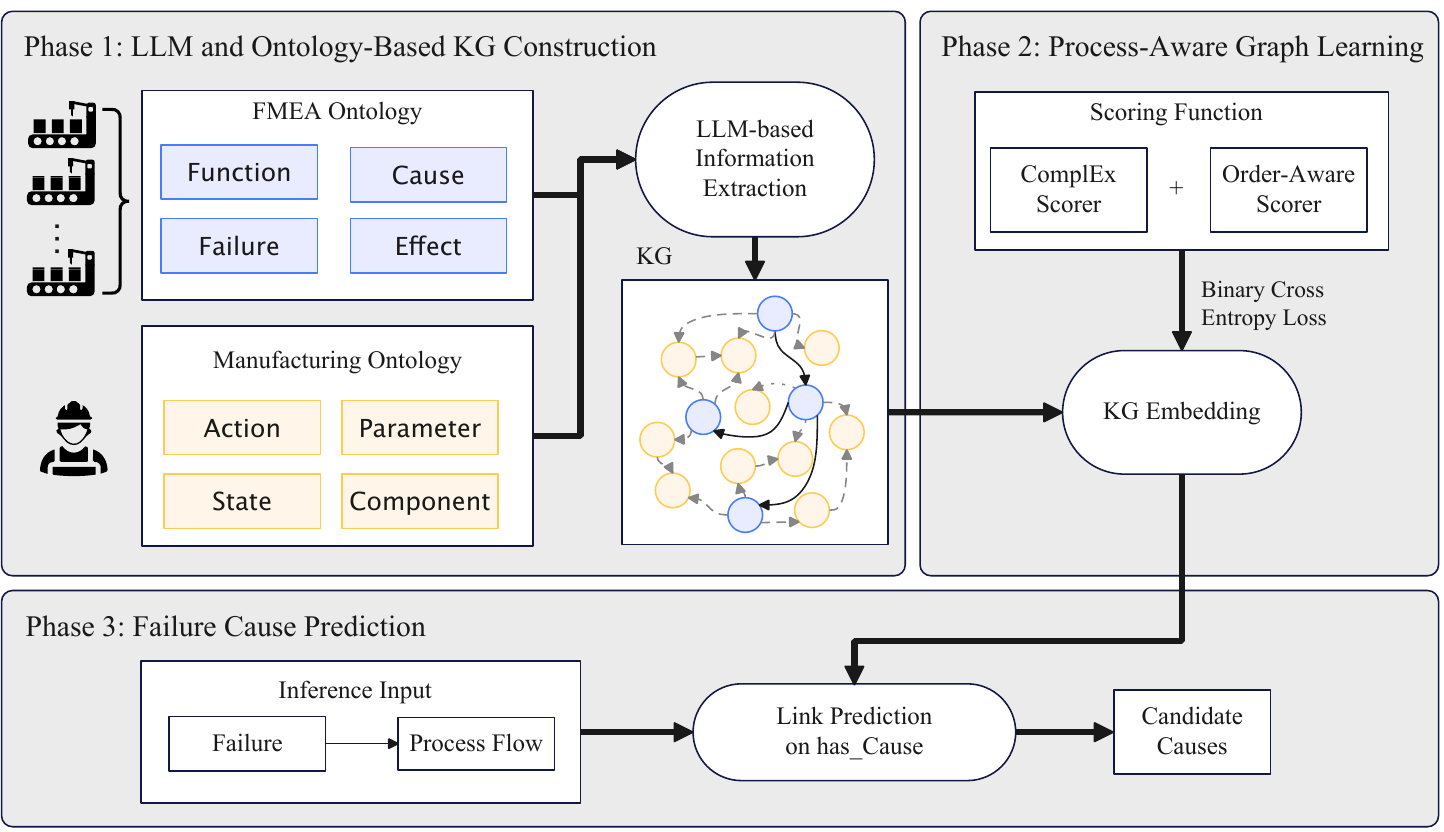} 
		\caption{Overall framework of OGPAL.}
		\label{fig:overview}
\end{figure}

As stated in the previous sections, OGPAL aims to enhance fault cause identification in manufacturing lines by leveraging manufacturing-domain conceptualization and process-aware graph learning. The overall framework is illustrated in Figure~\ref{fig:overview}, and it consists of three components. First, FMEA worksheets from multiple lines are processed by LLMs to extract entities such as actions, states, components, and parameters, which are aligned through a manufacturing-domain ontology and instantiated into a unified KG. This step replaces heavy manual ontology engineering and provides a generalizable representation of fault phenomena. Second, the KG is trained with an RGCN that learns embeddings reflecting semantic relationships and process consistency, so that elements which were not explicitly connected can be related through learned patterns. Finally, when a new failure occurs in a target line, its description and process context are used as input for link prediction, which ranks candidate causes in a way that remains consistent with the process flow. Sample code for the method is available at \url{https://github.com/syoka4156/ogpal.git}.

\subsection{Task Definition}

We formulate fault cause identification as a closed-set retrieval task over FMEA-derived cause candidates. Let $\mathcal{D}_{\mathrm{KB}}$ denote the reference FMEA records available for knowledge reuse, and let $G_{\mathrm{KB}}=(V,E,\mathcal{R})$ be the knowledge graph constructed from these records. Because worksheet entries of Failure, Cause, and Effect are represented using the same failure-related ontology concept, we define the candidate set by its provenance:

\[
\mathcal{C}_{\mathrm{cand}}
=
\{v\in V \mid v \text{ is instantiated from Failure, Cause, and Effect entries in }
\mathcal{D}_{\mathrm{KB}}\}.
\]

Given a query

\[
q=(x_q,\phi_q,\Pi_q),
\]

where $x_q$ is a failure description, $\phi_q$ is the process function at which the failure is observed, and $\Pi_q$ is the ordered process flow of the target line, the objective is to retrieve an ordered list of $n$ cause candidates:

\[
\mathcal L_n(q)
=
\operatorname{Top}_{n}
\left\{
\rho_{\Theta}(q,c\mid G_{\mathrm{KB}})
:
c\in\mathcal{C}_{\mathrm{cand}}
\right\},
\]

where $\rho_{\Theta}$ denotes the final relevance score assigned to candidate $c$. $\mathcal L_n(q)$ is evaluated against the expert-identified cause set $\mathcal E_q$.

\subsection{Knowledge Graph Construction}

The first step involves constructing a unified KG that integrates FMEA knowledge from multiple manufacturing lines. This process consists of three components: (i) ontology schema design, (ii) ontology-guided entity extraction using LLMs, and (iii) KG population that instantiates entities and relations. The ontology schema design (i) is described in Section~\ref{sec:onto_def}, while Section~\ref{sec:entity_extraction} presents the extraction (ii) and population (iii) processes.

\subsubsection{Ontology Definition}
\label{sec:onto_def}

\begin{figure}[!ht]
		\centering
		\includegraphics[width=\textwidth]{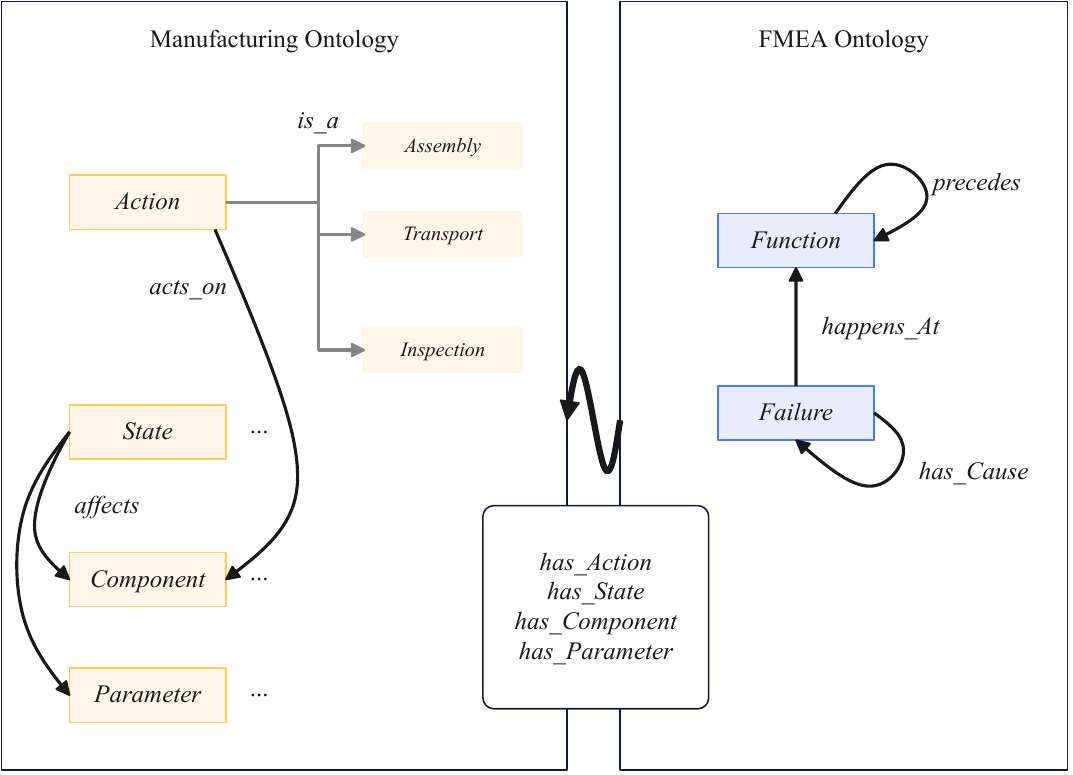} 
		\caption{Ontology description of manufacturing and FMEA concepts.}
		\label{fig:ontologies}
\end{figure}

This subsection describes step (i), where the ontology schema (definition of concepts and relations) is established. Figure~\ref{fig:ontologies} illustrates the core concepts and relationships defined in the manufacturing ontology and the FMEA ontology. The manufacturing ontology focuses on domain-level concepts such as \textit{Action}, \textit{State}, \textit{Component}, and \textit{Parameter}. These concepts are designed to be general enough to cover various manufacturing lines while being specific enough to capture essential fault phenomena. \textit{Action} denotes operations performed by equipment (e.g., assembly, bonding), \textit{State} represents conditions of components or human operators, including both nominal and degraded conditions (e.g., misalignment, contamination, incorrect teaching), \textit{Component} refers to physical entities involved in the process, including both discrete parts (e.g., robot, jig) and materials (e.g., gel, chip), and \textit{Parameter} covers measurable attributes (e.g., force, brightness). Among these, \textit{Action} and \textit{State} are emphasized because they directly capture how failures occur. They are connected to other concepts through relations such as \textit{acts\_on} and \textit{affects}. This representation is derived from recurring patterns observed in FMEA descriptions and can be interpreted as a lightweight abstraction of behaviour-state relationships in Umeda's FBS-based frameworks where behaviour is reflected through observable state transitions. 

While device-oriented functional ontologies distinguishes between device-centred and process-centred perspectives, failure phenomena in conventional FMEA descriptions may be expressed from either the operator or operand side. In device-oriented functional ontologies, a device plays an agent role by acting on an operand and changing its state \cite{kitamura2006ontological}. For example, failures may be described as an abnormal state of an operand (e.g., part misalignment) or as an incorrectly performed action by an operator (e.g., improper assembly). To ensure consistency in causal modelling, the proposed ontology integrates these perspectives into a unified representation. In manufacturing systems, equipment and human workers typically assume operator roles, while workpieces and materials assume operand roles. Process parameters are represented as attributes or conditions associated with operators, operands, or activities, rather than as operands themselves. Actions are treated as causal occurrences, whereas states may represent both preconditions and resulting conditions. This enables both types of descriptions to be mapped into a consistent structure, preserving expressiveness while maintaining causal coherence.

The FMEA ontology, on the other hand, structures the conventional worksheet elements by focusing on relationships among \textit{Failure}, \textit{Cause}, and \textit{Effect}, and their link to \textit{Function}. Worksheet entries of Failure, Cause, and Effect are abstracted into the \textit{Failure} concept and connected by \textit{has\_Cause}. The \textit{Function} concept is linked to \textit{Failure} through \textit{happens\_At}, which specifies the process step affected by a failure. It should be noted that \textit{Function} here follows the FMEA convention and represents process steps rather than design intent as defined in FBS frameworks. While FMEA ontologies in related works often contain solutions, RPNs, and other contextual information in addition to these four failure-related concepts \cite{mikos2011system,rehman2016ontology,camarillo2018knowledge}, we focus on the core concepts and their relationships to ensure generalizability across lines. Additionally, since functions in worksheets are often listed in process order, we additionally define the \textit{precedes} relation to encode sequential dependencies between functions. This relation ensures that causal reasoning respects the logical progression of manufacturing processes.

This design reflects the pragmatic structure of FMEA, where failures are described in relation to process functions and their causal chains. In this sense, failures are interpreted as state deviations of components and parameters, which is consistent with behaviour-based interpretations in Umeda's FBS, while avoiding the need for explicitly modelling full structures of FBS frameworks or causal role representations in ontologies such as MALFO. While such frameworks provide a more expressive representation, our simplified abstraction prioritizes compatibility with real-world FMEA data and robustness across heterogeneous manufacturing lines.

\subsubsection{Ontology-Guided LLM-Driven Entity Extraction}
\label{sec:entity_extraction}

This subsection describes steps (ii) and (iii), covering ontology-guided entity extraction and KG population. To populate the KG, FMEA worksheets are processed by LLMs guided by the manufacturing ontology. First, each worksheet sentence is parsed by the LLM with a structured prompt, which assigns candidate entities to the categories \textit{Action}, \textit{State}, \textit{Component}, and \textit{Parameter} (ii). Second, candidate strings are lightly normalized and compared with ontology labels using simple string-matching heuristics to produce a short list of identifiers. This aims for consistency with the ontology while allowing for variations in terminology such as ``place'' and ``placement''. The LLM is prompted again to select the most appropriate identifier from the shortlist or to create a new entry when no candidate fits (iii). This two-stage prompting ensures that extracted terms remain consistent with the ontology while allowing flexibility to capture unseen expressions.

For example, given ``Placing a chip'', the LLM extracts (Action: placement, Component: chip). These are matched to ontology candidates (e.g., [``place'', ``placer'', ``pallet'']), from which the LLM selects \{``element'': ``place''\}. If no candidate is suitable, a new entry is created under an appropriate concept (e.g., ``chip'' under ``electric part''). The selected identifiers are then instantiated in the KG, resulting in relations such as place \textit{acts\_on} chip. This process is repeated for each sentence, enabling structured FMEA knowledge to be integrated across lines.

\begin{figure}[!ht]
	\centering
	\includegraphics[width=\textwidth]{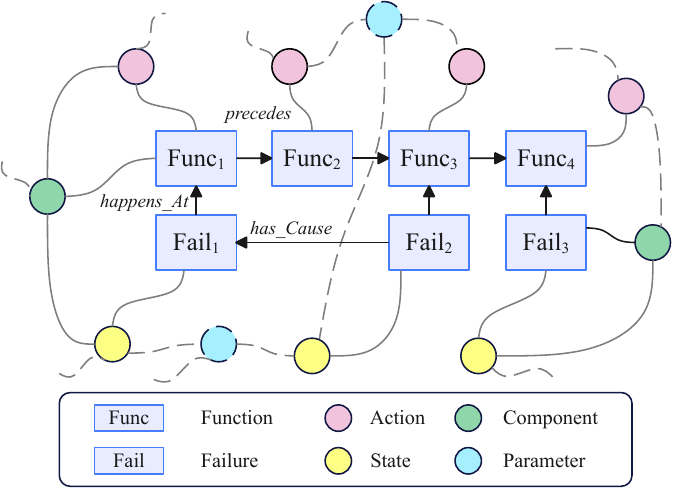} 
	\caption{Overview of the constructed knowledge graph. Instances from FMEA ontology and manufacturing ontology are linked through defined relations. Solid lines represent elements directly extracted from descriptions, while dashed lines represent those connected in the KG.}
	\label{fig:kg}
\end{figure}

The resulting entities and their relations are instantiated as nodes and edges in a unified KG (Figure~\ref{fig:kg}). Relations such as \textit{acts\_on}, \textit{affects}, \textit{has\_Cause}, \textit{happens\_At}, and \textit{precedes} encode semantic and process dependencies. For example, a \textit{Function} such as ``chip placement'' can be represented as an \textit{Action} ``place'' that acts on a \textit{Component} ``chip'' ($\text{Func}_1$), while a \textit{Failure} such as ``chip misalignment'' can be represented as a \textit{State} ``misalignment'' that affects the same component ($\text{Fail}_1$). Even if an FMEA instance cannot be directly mapped to multiple concepts ($\text{Func}_4$), it can be linked to other concepts through edges extracted from other descriptions. In this way, FMEA concepts, such as \textit{Function} and \textit{Failure}, from different lines can be connected through shared manufacturing-domain concepts including \textit{Action}, \textit{State}, \textit{Component}, and \textit{Parameter}, allowing the model to learn from commonalities across heterogeneous descriptions. It should be noted that while well-established design frameworks including FBS models define functions associated with components and their behaviours, OGPAL focuses on the practical structure of FMEA descriptions, where functions are not always explicitly linked to components. Instead, the proposed KG allows for flexible connections between functions and other concepts, enabling the representation of a wider range of failure phenomena as observed in real-world FMEA records. The pseudo-code of the extraction algorithm and the LLM prompts are provided in \ref{sec:appendix_entity_extraction}.

\subsection{Process-Aware Graph Learning}

\begin{figure}[!ht]
	\centering
	\includegraphics[width=\textwidth]{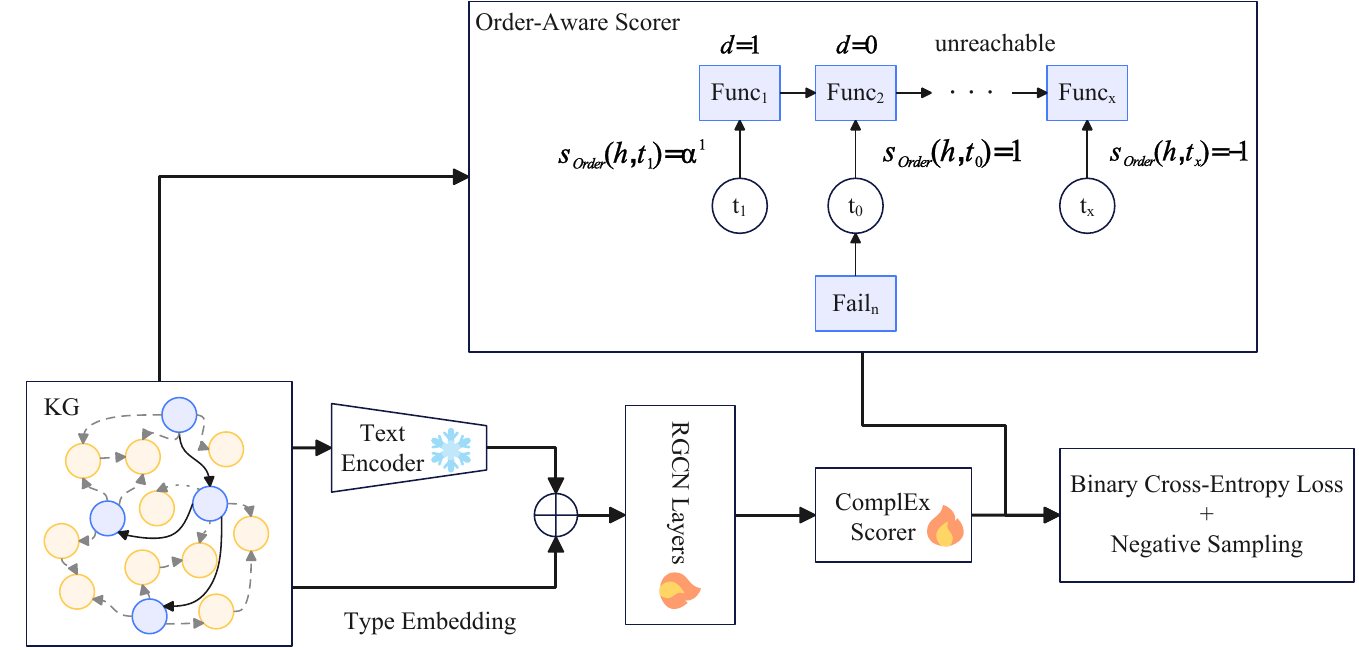} 
	\caption{Architecture overview of process-aware graph learning.}
	\label{fig:arch}
\end{figure}

In the previous section, we constructed a unified KG that integrates FMEA knowledge from multiple manufacturing lines together with manufacturing-domain concepts. The next step is to train a GNN on this KG in order to learn embeddings that capture semantic relationships while also remaining consistent with process flows. 

We adopt an RGCN \citep{schlichtkrull2018modeling} as the base encoder, since it is well-suited for multi-relational graphs derived from heterogeneous knowledge sources. RGCN propagates and aggregates information across relation-specific neighborhoods, producing node embeddings that reflect the semantics of actions, states, components, and parameters as defined in the ontology. For link prediction, we adopt the ComplEx scoring function \citep{trouillon2016complex}, 
which has proven effective for modeling asymmetric relations such as causal dependencies. Given a triplet $(h, r, t)$, where $h$ and $t$ denote the head (subject) and tail (object) entities 
and $r$ denotes the relation, the score is computed as follows:

\begin{equation}
\label{eq:complex}
s_{\text{ComplEx}}(h, r, t) = \mathrm{Re}\,\langle \mathbf{e}_h, \mathbf{e}_r, \overline{\mathbf{e}_t} \rangle,
\end{equation}

\noindent
where $\mathbf{e}_h$ and $\mathbf{e}_t$ are the complex-valued embeddings of the head and tail entities, $\mathbf{e}_r$ is the relation embedding, and $\overline{\mathbf{e}_t}$ denotes the complex conjugate of $\mathbf{e}_t$. This formulation captures both symmetric and asymmetric relational patterns, which is essential for representing causal links in manufacturing processes.

Although semantic relationships are learned through ontology-guided embeddings, causal reasoning must also respect the sequential flow of manufacturing processes. To this end, we introduce an order-aware scoring function into the training objective. For each triplet, the head and tail failure entities are first mapped to their associated manufacturing processes through the \textit{happens\_At} relation. We then define the distance \(d(h,t)\) between the corresponding processes based on their relative order in the process flow:

\begin{equation}
	d(h, t) =  
	\begin{cases}
		0 & \text{if $h$ overlaps with $t$} \\
		k & \text{if $h$ precedes $t$} \\
		\infty & \text{otherwise}
	\end{cases}
\end{equation}

\noindent
where $k$ is a constant number of steps between processes. The distance $d(h,t)$ captures whether the head process occurs before, during, or after the tail process. We then define the order-aware scoring function as follows:

\begin{equation}
	s_{order}(d) = 
	\begin{cases}
		1 & \text{if $d = 0$} \\
		\alpha^d & \text{if $d \geq 1$} \\
		-1 & \text{if $d = \infty$} \\
		\beta & \text{if unknown}
	\end{cases}
\end{equation}

\noindent
where $\alpha$ is a decay factor, and $\beta$ is a penalty weight for unknown relations in $happens\_At$. The function $s_{order}(d)$ encourages the model to assign higher scores to triplets that conform to process order and penalizes those that contradict it. This ensures that predictions remain consistent with the logical progression of manufacturing operations.

The overall training objective combines the ComplEx scoring function with the order-aware component. The final score for a triplet is given by:

\begin{equation}
	s(h,r,t) = s_{ComplEx} (h,r,t) + \lambda s_{order}(d(h,t))
\end{equation}

\noindent
where $\lambda$ is a hyperparameter that balances the influence of process order. The model is trained using binary cross-entropy loss with logits. This objective encourages high scores for observed (positive) triplets and low scores for artificially generated negative triplets. The negative samples are obtained by randomly corrupting the tail entity for each positive triplet.

In this way, the model jointly optimizes semantic consistency from the ontology and relational structure together with process consistency derived from manufacturing flows. The resulting embeddings form the basis for predicting fault causes in a manner that is both transferable across heterogeneous lines and coherent with process order.

\subsection{Fault Cause Retrieval}

\begin{figure}[!ht]
	\centering
	\includegraphics[width=\textwidth]{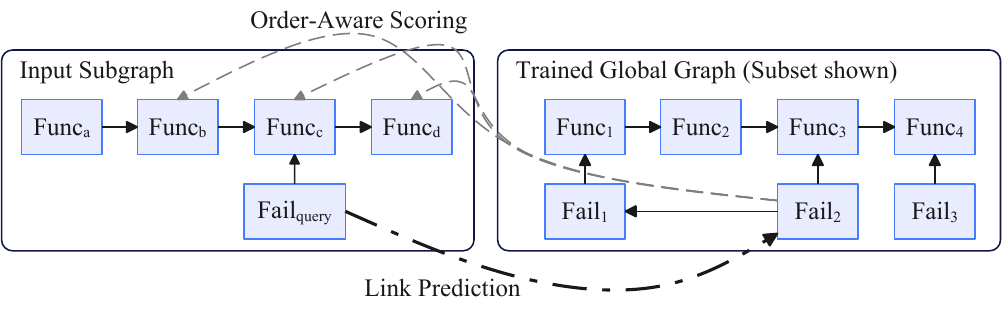} 
	\caption{Inference Process. Descriptions of new failures and their process contexts are input to the trained graph-based scorer to predict and rank candidate causes.}
	\label{fig:inference}
\end{figure}

The final step is to retrieve and rank fault causes using the trained graph-based scorer. Figure~\ref{fig:inference} summarizes the inference flow. When a new failure occurs, its description together with the corresponding process function is processed by the same LLM-based extraction method introduced in Section~\ref{sec:entity_extraction}. The extracted entities are then mapped to nodes in the KG. The trained graph-based scorer performs link prediction on these nodes, producing scores for candidate causes through the \textit{has\_Cause} relation. Candidate causes are ranked according to their scores, and the top-$n$ results are returned as potential explanations. A final re-ranking step considers process flow information so that causes linked to processes occurring before the failure's process are prioritized. This ensures that the inferred causes are not only semantically relevant but also temporally plausible within the manufacturing context. In this way, the framework provides operators with a set of consistent and transferable candidate causes, supporting diagnosis across heterogeneous manufacturing lines.

\section{Case Study}
\label{sec:case}

\subsection{Case Study Design}

This case study evaluates OGPAL through three complementary assessments in the context of automotive pressure sensor assembly lines. The assessments are designed to systematically evaluate the accuracy of LLM-driven entity extraction, the accuracy of fault cause prediction, and the contributions of individual components through ablation studies.

\paragraph{Assessment 1: Accuracy of LLM-driven entity extraction} We evaluate whether the LLM correctly extracts \textit{Action}, \textit{State}, \textit{Component}, and \textit{Parameter} entities from FMEA records according to the manufacturing ontology. For each of the \textit{Failure} and \textit{Function} entry types, 50 records are randomly sampled from the constructed KG and evaluated against manual annotations using the F1 score. This assessment validates the quality of the KG construction step before proceeding to inference.

\paragraph{Assessment 2: Fault cause retrieval against expert-identified cause sets} We evaluate the ability of OGPAL to rank valid fault causes for three fault scenarios defined in a real automotive pressure sensor manufacturing line. Ground-truth causes for each scenario were identified through interviews with two expert operators. Performance is measured using Precision, Recall, F1 score at n (P@$n$, R@$n$, F1@$n$), and nDCG. The framework is compared against two baselines: a RAG-based approach \citep{bahr2025knowledge}, which relies on linguistic similarity without structured graph reasoning, and an RGCN-based model \citep{trouillon2016complex}, which shares a similar graph learning architecture but does not incorporate domain conceptualization or process-aware scoring. The RAG-based approach represents recent methods that leverage LLMs and retrieval mechanisms to reuse FMEA knowledge based on semantic similarity, while the RGCN-based model serves as a representative graph learning baseline that captures structural dependencies without domain-specific constraints. These baselines are selected to disentangle the contributions of linguistic retrieval, graph-based reasoning, and domain- and process-aware modeling, enabling a systematic evaluation of OGPAL.

\paragraph{Assessment 3: Ablation study of individual components} To isolate the contributions of manufacturing-domain conceptualization, process-aware graph learning, and heterogeneous cross-line FMEA knowledge, we compare the full model with variants that remove the domain conceptualization, disable process-aware graph learning, or restrict the training data to FMEA records from the target line. The last comparison quantifies the extent to which knowledge transferred from other manufacturing lines improves fault-cause identification on the target line. Performance is measured using Precision, Recall, F1 score, and nDCG on the same three fault scenarios.

The evaluation draws on two types of data: a heterogeneous collection of FMEA records from multiple manufacturing lines used for model training, and process flow and failure data from a target line used for inference and evaluation. The training data include FMEA records from the target line because, in the intended operational setting, these records are developed during the line-design phase and are therefore available before deployment. However, the evaluation scenarios and the ground-truth cause sets are excluded from knowledge-graph construction and the model-training. Further details of the data are provided in Section~\ref{sec:data_description}, followed by the evaluation metrics in Section~\ref{sec:evaluation_metrics}.

\subsection{Data Description}
\label{sec:data_description}
In total, the collected datasets contain 1,213 records across 11 FMEA datasets covering diverse manufacturing processes. These include visual inspection, chip- and board-level assembly, adhesive processes, workpiece transfer, die-casting burr removal, machining operations, body assembly and packing, and radar module assembly. The products involved range from small electronic components to large automotive parts, such as commutators, pressure sensors, and automotive body components, and the equipment includes robots, conveyors, vision systems, and various specialized machinery. Each FMEA record (Function, Failure, Cause, Effect) was decomposed into four domain concepts (\textit{Action}, \textit{State}, \textit{Component}, \textit{Parameter}) and mapped to the KG (see Section~\ref{sec:entity_extraction}). This step operationalizes the manufacturing-domain conceptualization introduced earlier. We used the LLM \textit{gpt-4o-2024-11-20} \citep{hurst2024gpt} to extract entities and relationships, resulting in a unified KG with 4,253 nodes and 9,341 edges. The manufacturing ontology contained 44 \textit{Action}, 16 \textit{State}, 23 \textit{Component}, and 14 \textit{Parameter} classes. All records were originally written in Japanese and processed in the original language without manual translation.

\begin{figure}[!ht]
	\centering
	\includegraphics[width=0.6\textwidth]{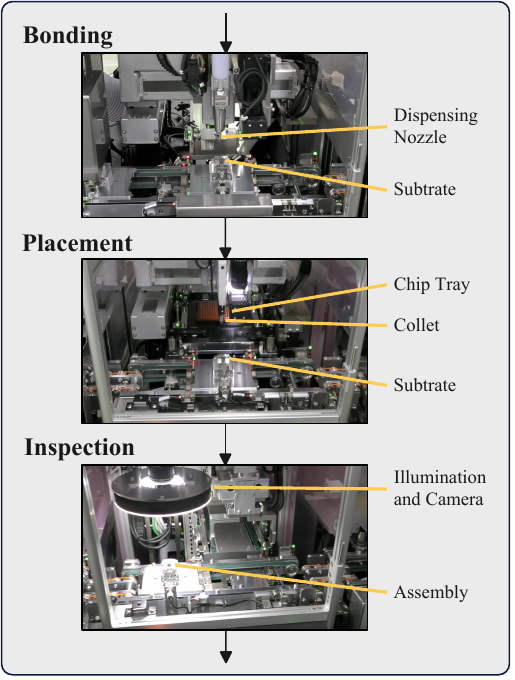} 
	\caption{Processes in the target manufacturing line. Images of bonding, placement, and visual inspection processes are shown from top to bottom. These processes are sequentially connected to other processes to form the entire manufacturing line.}
	\label{fig:target_line}
\end{figure}

The target system for inference is an automated manufacturing line for automotive pressure sensors, including chip assembly, adhesive application, and visual inspection. Figure~\ref{fig:target_line} illustrates the chip assembly process, where a robot picks up a chip from a tray and places it onto a Subtrate. The process involves actions such as moving, gripping, and releasing; components such as the robot arm, gripper, and vision system; and parameters such as position coordinates and gripping force.

To evaluate fault cause retrieval, we interviewed two expert operators with extensive experience in diagnosing this line. They identified three common failure scenarios: (i) misalignment of the chip during placement, leading to assembly defects; (ii) grasping failure of the chip, resulting in unsuccessful pickup; and (iii) false detection during visual inspection, causing good parts to be rejected. Since these scenarios are closely tied to the sequence of operations, they also highlight the importance of incorporating process flow into causal reasoning.

For each query scenario \(q\), the experts produced a list of plausible causes based on their operational experience. The cause sets were prepared independently of the FMEA records and were not used for knowledge graph construction, model training, prompt development, or model selection. We denote the resulting expert-elicited reference set by $\mathcal{E}_q$. The expert lists were subsequently cross-checked to construct the final reference set used for evaluation. All methods were evaluated on the same set of failure scenarios, with the corresponding cause set for each scenario used as ground truth.

\subsection{Experimental Setup}
\label{sec:experimental_setup}

We trained a two-layer RGCN with a hidden size of 128. Node features combined 128-dimensional PCA-reduced text embeddings with 16-dimensional type embeddings. The model was optimized with AdamW (learning rate $1\times10^{-3}$) for 1,000 epochs using binary cross-entropy with 1:5 negative sampling. For the order-aware scoring function, we set $\alpha=0.7$, $\beta=-0.1$, and $\lambda=0.6$. All experiments were conducted on a single NVIDIA A100 80GB GPU with a fixed random seed. 

The R-GCN was trained on a randomly selected subset of the positive \textit{has\_Cause} triples, with a disjoint subset used for validation. Final performance was evaluated exclusively using three separately constructed fault-scenario queries. All instances are treated uniformly, without incorporating severity measures such as RPN to avoid biasing the model towards more severe cases. 

We compare OGPAL with two baselines. The first baseline is the RAG-based approach \citep{bahr2025knowledge}. The retriever selects the 20 most similar \textit{Failure} entries from the FMEA KG using cosine similarity, and the LLM then generates candidate causes based on these retrieved entries. The second baseline is the RGCN-based link prediction model \citep{trouillon2016complex}, which uses the same architecture and training procedure as OGPAL but is trained directly on worksheet-level elements without manufacturing-domain conceptualization or the order-aware scoring function. To ensure a fair comparison, the same \textit{text-embedding-3-small} model is used to generate the original text embeddings for all methods. In addition, the same completion model, \textit{gpt-4o-2024-11-20} \citep{hurst2024gpt}, is used wherever LLM-based generation is involved. The temperature of all LLM models is fixed to 0 to support reproducible generation.

\subsection{Evaluation Metrics}
\label{sec:evaluation_metrics}

We evaluate ranking performance using precision, recall, and F1 score at cutoff \(n\) (P@\(n\), R@\(n\), and F1@\(n\), respectively). Because the same cause may appear more than once in a ranked list, each ground-truth cause is counted as correct only at its first occurrence. We therefore define the unique-relevance indicator as

\begin{equation}
\label{eq:unique_relevance}
r_{q,i}
=
\mathbf{1}
\left[
c_{q,i} \in \mathcal{E}_q
\;\land\;
c_{q,i} \notin \{c_{q,1},\ldots,c_{q,i-1}\}
\right],
\end{equation}

where \(\mathbf{1}[\cdot]\) returns 1 if the condition is true and 0 otherwise. Predictions that do not correspond to a ground-truth cause, including missing predictions, receive \(r_{q,i}=0\).

Using this indicator, Precision, Recall, and F1 score at cutoff \(n\) for scenario \(q\) are defined as

\begin{equation}
\label{eq:precision}
\mathrm{P@}n(q)
=
\frac{1}{n}
\sum_{i=1}^{n} r_{q,i},
\end{equation}

\begin{equation}
\label{eq:recall}
\mathrm{R@}n(q)
=
\frac{1}{|\mathcal{E}_q|}
\sum_{i=1}^{n} r_{q,i},
\end{equation}

and

\begin{equation}
\label{eq:f1}
\mathrm{F1@}n(q)
=
\begin{cases}
\displaystyle
\frac{
2\,\mathrm{P@}n(q)\,\mathrm{R@}n(q)
}{
\mathrm{P@}n(q)+\mathrm{R@}n(q)
},
&
\mathrm{P@}n(q)+\mathrm{R@}n(q)>0,\\[8pt]
0, & \text{otherwise}.
\end{cases}
\end{equation}

To aggregate the cutoff-specific results across scenarios, we compute an equally weighted macro average for each metric, P@\(n\), R@\(n\), and F1@\(n\). In the evaluated scenarios, the number of ground-truth causes, \(|\mathcal{E}_q|\), ranges from 10 to 23. We evaluate the cutoff-specific metrics for \(n=1,\ldots,20\) to examine performance across different diagnostic depths.

% In addition, we use average precision at rank 20 (AP@20) to summarize the quality of the complete evaluated ranking. AP@20 rewards methods that retrieve a larger number of unique ground-truth causes and rank them earlier. For each scenario \(q\), it is defined as

% \begin{equation}
% \label{eq:ap20}
% \mathrm{AP@20}(q)
% =
% \frac{1}{|\mathcal{E}_q|}
% \sum_{i=1}^{20}
% \mathrm{P@}i(q)\,r_{q,i}.
% \end{equation}

% Thus, only ranks at which a previously unseen ground-truth cause is retrieved contribute to AP@20. Ground-truth causes not retrieved within the top 20 make no contribution. Finally, we report mean average precision at rank 20 (mAP@20), calculated as the equally weighted mean of AP@20 across all failure scenarios.

In addition, we use normalized discounted cumulative gain at rank 20 (nDCG@20) to summarize the quality of the complete evaluated ranking. nDCG@20 rewards methods that retrieve unique ground-truth causes at higher ranks. For each scenario \(q\), it is defined as

\begin{equation}
\label{eq:ndcg20}
\begin{aligned}
\mathrm{DCG@20}(q)
&=
\sum_{i=1}^{20}
\frac{r_{q,i}}{\log_2(i+1)},\\
\mathrm{nDCG@20}(q)
&=
\frac{\mathrm{DCG@20}(q)}
{\mathrm{DCG@20}_{\max}(q)},
\end{aligned}
\end{equation}

\noindent
where \(\mathrm{DCG@20}_{\max}(q)\) is the maximum possible DCG@20 for scenario \(q\). We also report the macro-average of nDCG@20 across all failure scenarios.

\subsection{Assessment 1: Accuracy of LLM-Driven Entity Extraction}
\label{sec:result_entity_extraction}

Before evaluating fault cause prediction, we first assess the quality of the ontology-guided LLM-driven entity extraction described in Section~\ref{sec:entity_extraction}. For each Failure and Function entry, we randomly sampled 50 records from the constructed KG and evaluated the accuracy of extracted sets of \textit{Action}, \textit{State}, \textit{Component}, and \textit{Parameter} entities by matching them against annotations. This design allows partial correctness to be accumulated across instances, reflecting the actual usability of extracted entities for knowledge graph construction.

\begin{table}[!ht]
\centering
\caption{F1 score of ontology-guided LLM-driven entity extraction.}
\label{tab:entity_extraction_accuracy}
\begin{tabular}{lcccc}
	\toprule
         & Action & State & Component & Parameter \\
	\midrule
Failure  & 0.941  & 0.989 & 0.957     & 0.857     \\
Function & 0.958  & 1.000 & 0.973     & 0.791    
\\
	\bottomrule
\end{tabular}
\end{table}

Table~\ref{tab:entity_extraction_accuracy} shows the F1 scores for each entity type extracted using the ontology-guided LLM-driven approach. Overall, the proposed extraction method achieves consistently high F1 scores for \textit{Action}, \textit{State}, and \textit{Component}, indicating that operational activities, physical elements, and fault conditions are effectively captured. While \textit{State} entities from Function entries achieve perfect scores, the number of such entries is limited (4 instances), which may have contributed to this result. In contrast, those from Failure entries are more frequent (45 instances) and still achieve a high F1 score of 0.989, demonstrating the method's robustness. 

\textit{Parameter} entities show slightly lower performance, particularly for Function entries (F1 = 0.791), which can be attributed to the abstract and context-dependent nature of parameter descriptions in FMEA worksheets. For example, in the function description ``connector inspection by camera'', the term ``connector'' was incorrectly classified as \textit{Parameter}, although it actually denotes the inspection target component, reflecting semantic ambiguity in concise FMEA descriptions. 

\subsection{Assessment 2: Fault Cause Prediction against Expert Ground Truth}
\label{sec:result_fmea}

\begin{figure}[!ht]
	\centering
	\includegraphics[width=\textwidth]{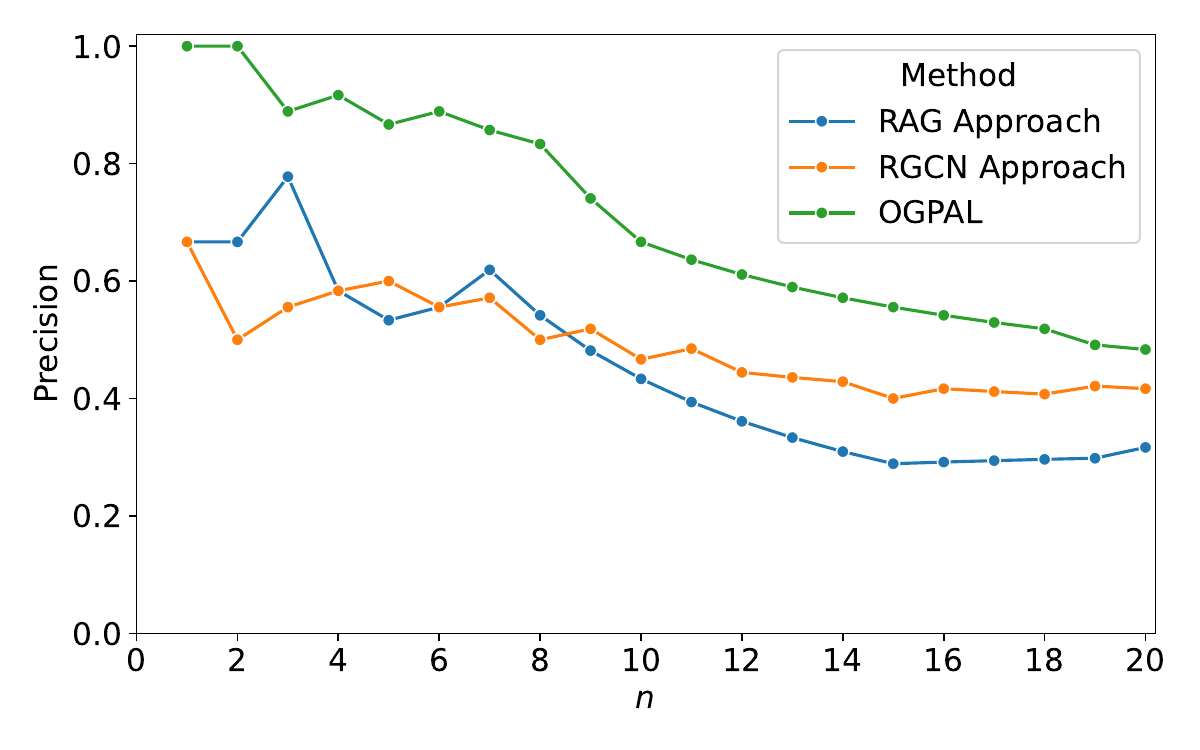} 
	\caption{Precision at n (P@$n$) comparison between OGPAL and baselines}
	\label{fig:pr_comparison}
\end{figure}

\begin{figure}[!ht]
	\centering
	\includegraphics[width=\textwidth]{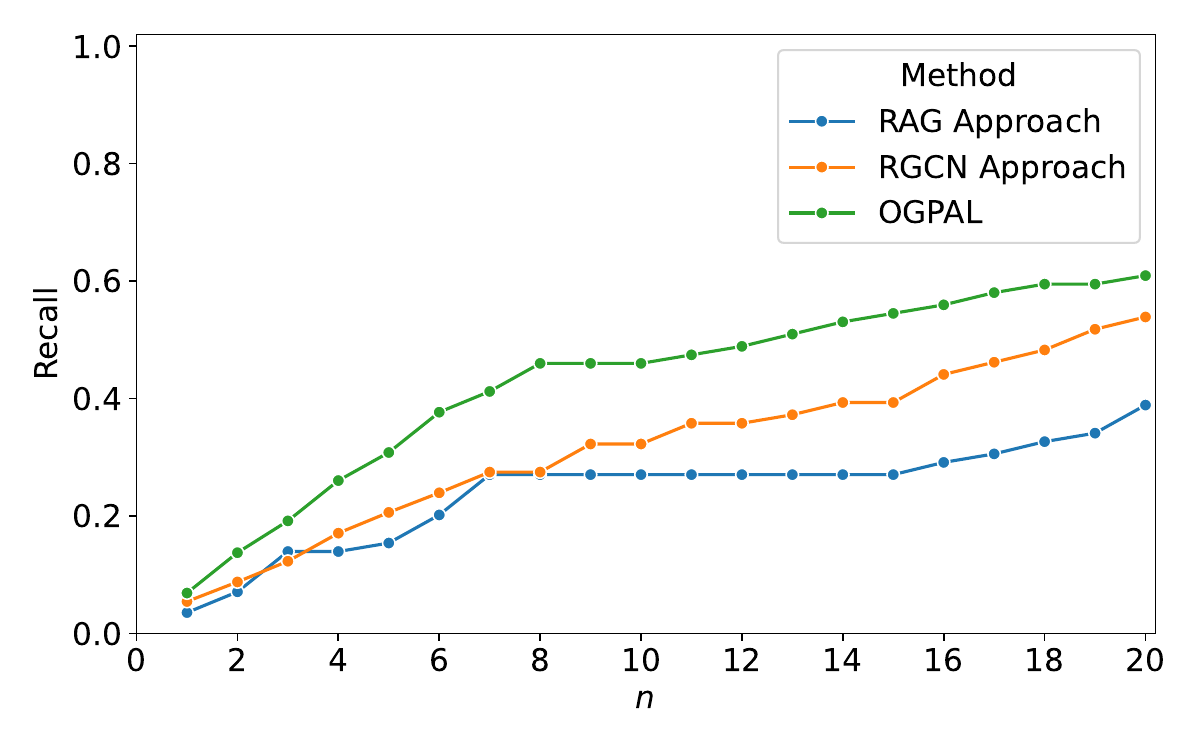} 
	\caption{Recall at n (R@$n$) comparison between OGPAL and baselines}
	\label{fig:rc_comparison}
\end{figure}

\begin{figure}[!ht]
	\centering
	\includegraphics[width=\textwidth]{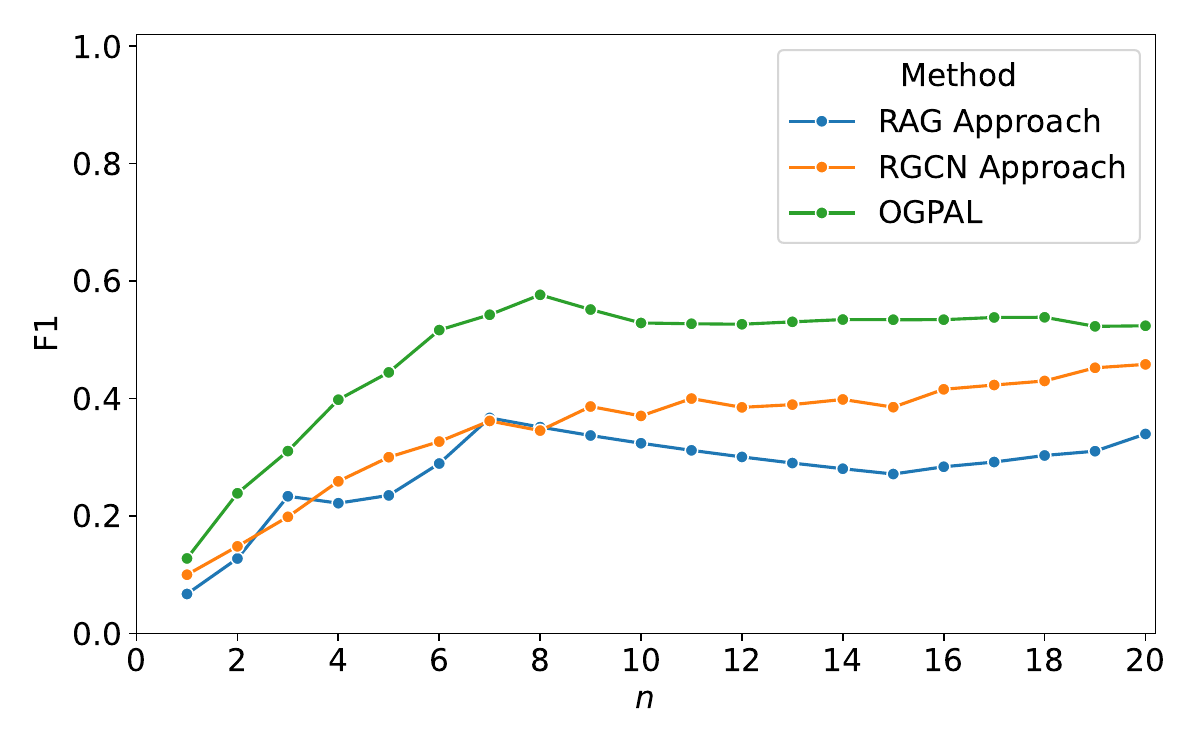} 
	\caption{F1 score comparison between OGPAL and baselines}
	\label{fig:f1_comparison}
\end{figure}

Figures~\ref{fig:pr_comparison}-\ref{fig:f1_comparison} compare the P@$n$, R@$n$, and F1@$n$ with the baselines. Higher precision indicates that the inferred candidates are more likely to be valid causes, while higher recall indicates that a larger proportion of ground-truth causes are retrieved. The F1 score balances these two aspects. OGPAL consistently outperformed both the RAG- and RGCN-based approaches across all evaluation metrics and cutoff values. Its advantage was particularly pronounced at more restrictive cutoffs: it achieved a P@10 of 0.667, compared with 0.433 for RAG and 0.467 for RGCN. It also obtained an R@10 of 0.460 and an F1@10 of 0.529, substantially exceeding the corresponding baseline scores. These results indicate that OGPAL not only retrieves a broader set of ground-truth causes but, more importantly, ranks valid causes more densely within the top portion of the candidate list. At $n=20$, the framework maintained its overall advantage, achieving a P@20 of 0.483, an R@20 of 0.609, and an F1@20 of 0.524.

\begin{table}[!ht]
\centering
\caption{Performance comparison of nDCG@20 and its macro-average. Best results are highlighted in bold.}
\label{tab:ndcg20}
\begin{tabular}{lcccc}
\toprule
Method & Scenario 1 & Scenario 2 & Scenario 3 & Average \\
\midrule
RAG Approach   & 0.555     & 0.482       & 0.312      & 0.450     \\
RGCN Approach   & 0.422      & 0.604     & 0.652      & 0.559      \\
OGPAL   & \textbf{0.705}      & \textbf{0.657}      & \textbf{0.797}      & \textbf{0.719} \\
\bottomrule     
\end{tabular}
\end{table}

Table~\ref{tab:ndcg20} further supports the superior ranking capability of OGPAL by reporting the scenario-level nDCG@20 and its macro-average. The framework achieved the highest nDCG@20 in all three scenarios, yielding a macro-average of 0.719, compared with 0.450 for RAG and 0.559 for RGCN. The improvement was particularly substantial in Scenario~3, where OGPAL achieved an nDCG@20 of 0.797, compared with 0.312 for RAG and 0.652 for RGCN. Because nDCG@20 accounts for the positions of relevant candidates within the ranked list, these results indicate that the framework consistently places valid causes closer to the top across different scenarios. This finding complements the strong P@10 results and provides further evidence of its early-ranking effectiveness.

\begin{table}[!ht]
\centering
\caption{Top five outputs of each method for the false detection during visual inspection (English translation).}
\label{tab:example_outputs}
\begin{tabular}{lll}
\toprule
\textbf{Method} & \textbf{Predicted Causes} \\ 
\midrule
RAG Approach&
\begin{tabular}[t]{@{}l@{}}
False detection \\
Misclassification in manual inspection mode \\
Insufficient inspection accuracy \\
Laser misjudgment \\
Incorrect selection of inspection program
\end{tabular} \\[3pt]
\midrule

RGCN Approach&
\begin{tabular}[t]{@{}l@{}}
Position misalignment \\
Vibration \\
Product position offset \\
Memory offset \\
Positional misalignment
\end{tabular} \\[3pt]
\midrule

OGPAL &
\begin{tabular}[t]{@{}l@{}}
Memory offset \\
Wrong Subtrate type \\
Vibration \\
Camera misalignment \\
Foreign particle adhesion
\end{tabular} \\

\bottomrule
\end{tabular}
\end{table}

Table~\ref{tab:example_outputs} presents the top five predicted causes for the ``false detection during visual inspection'' scenario from each method. OGPAL successfully identified relevant causes such as ``Memory offset'' and ``Wrong Subtrate type,'' which are directly related to the visual inspection process. In contrast, the RAG-based method produced more generic and linguistically similar causes such as ``false detection'' which provide limited insights for diagnosis. The RGCN baseline suggested overly broad causes such as ``Position misalignment'', reflecting its lack of process awareness and overreliance on local co-occurrence patterns in the KG.

The advantage over the RGCN baseline is evident in the early stage of the ranking, which is crucial for practical diagnosis. The baseline mistakenly ranked causes related to ``improper equipment grounding'' as the top cause for the ``grasping failure of the chip'' scenario, which is not related to the target manufacturing line. This indicates that without process-aware graph learning, the model can produce inconsistent predictions that violate process flow, and without manufacturing-domain conceptualization, it can fail to generalize fault phenomena described differently in other lines.

The superiority over the RAG-based method shows that linguistic similarity alone cannot capture manufacturing semantics where similar terms may refer to different concepts. For example, as mentioned earlier, in the ``false detection during visual inspection'' scenario, the RAG-based method retrieved ``False detection'' as the highest-ranked cause, which is too generic and does not provide actionable insights. In addition, it retrieved ``Overvoltage stress'' as a candidate cause. This is related to the chip itself, but irrelevant to the visual inspection process. 

In addition, we report the inference latency to provide an indication of practical applicability. The reported values are averaged across 30 queries. The RAG baseline required approximately 0.91 seconds per query. In contrast, OGPAL required 10.34 seconds for knowledge graph construction from input text and 3.25 seconds for graph-based inference. The framework incurs higher latency primarily due to the additional steps of entity extraction, KG population, and graph-based inference. Since diagnosis in manufacturing contexts is often performed offline, this latency is acceptable for practical use. However, further optimization of the entity extraction and graph inference steps could reduce latency for real-time applications.

\subsection{Assessment 3: Ablation Study of Individual Components}
\label{sec:result_ablation}

\begin{table}[!ht]
\centering
\caption{Ablation study of OGPAL. H: Heterogeneous data, C: Conceptualization, P: Process-aware. The table reports Precision (P), Recall (R), F1, and nDCG at different cutoffs. Best results are highlighted in bold.}
\begin{tabular}{lccccc}
\toprule
Method & H/C/P & Metric & @1 & @10 & @20 \\
\midrule
\multirow{4}{*}{RGCN Baseline} & \multirow{4}{*}{\checkmark/--/--} 
 & P  & 0.667 & 0.467 & 0.417 \\
   &       & R  & --    & 0.322 & 0.538 \\
  &       & F1 & --    & 0.370 & 0.458 \\
  &       & nDCG & --    & 0.506 & 0.559 \\
\midrule
\multirow{4}{*}{%
  \shortstack[l]{OGPAL\\
                 w/o Process-Aware Learning}}
 & \multirow{4}{*}{\checkmark/\checkmark/--}
 & P  & \textbf{1.000} & \textbf{0.667} & 0.417 \\
    &       & R  & --    & \textbf{0.460} & 0.564 \\
    &       & F1 & --    & \textbf{0.529} & 0.465 \\
	&       & nDCG & --    & 0.747 & 0.669 \\
\midrule
\multirow{4}{*}{%
  \shortstack[l]{OGPAL\\
                 w/o Heterogeneous Data}}
 & \multirow{4}{*}{--/\checkmark/\checkmark}
 & P  & \textbf{1.000} & 0.533 & 0.383 \\
    &       & R  & --    & 0.358 & 0.503 \\
    &       & F1 & --    & 0.416 & 0.422 \\
	&       & nDCG & --    & 0.644 & 0.610 \\
\midrule
\multirow{4}{*}{OGPAL (Full)} 
 & \multirow{4}{*}{\checkmark/\checkmark/\checkmark}
 & P  & \textbf{1.000} & \textbf{0.667} & \textbf{0.483} \\
    &       & R  & --    & \textbf{0.460} & \textbf{0.609} \\
    &       & F1 & --    & \textbf{0.529} & \textbf{0.524} \\
    &       & nDCG & --    & \textbf{0.757} & \textbf{0.719} \\
\bottomrule
\end{tabular}
\label{tab:ablation}
\end{table}

Table~\ref{tab:ablation} presents the results of the ablation studies. H stands for training on heterogeneous data from multiple lines. C stands for manufacturing-domain conceptualization, and P stands for process-aware graph learning. When heterogeneous data are not used, the model is trained only on FMEA data from the target line. When conceptualization is not used, the concept-level ontology constraints are removed, and the model is trained on worksheet-level elements without mapping to domain concepts. When process-aware learning is not used, the order-aware scoring function is removed from the training objective. The first row shows the performance of the model without both manufacturing-domain conceptualization and process-aware graph learning. The second row shows the performance without process-aware graph learning. The third row shows the performance without heterogeneous data, meaning that the model was trained only on FMEA data from the target line. Finally, the last row presents the full proposed method, OGPAL. The results indicate that all components contribute to improved performance. In particular, the full model achieved a P@20 of 0.483, an R@20 of 0.609, an F1@20 of 0.524, and an nDCG@20 of 0.719, outperforming all ablated versions. In some cases, the ablated versions achieved the same metric at @1 and @10, but at @20, the full model consistently outperformed them. This indicates that the full model is more effective in retrieving a comprehensive set of valid causes, which is important for practical diagnosis. When the model is trained only on the target line without heterogeneous data, the performance drops significantly. This highlights the performance of leveraging diverse FMEA data to capture a wide range of fault phenomena.

Despite these strengths, our method also has limitations. For instance, in the ``misalignment of the chip during placement'' scenario, it mistakenly ranked ``camera misalignment'' as the top cause. This generally relates to misalignment, but no camera is involved in the target process step. This error likely stems from the limited representation of the target process flow which did not include the precise components involved. 

Overall, these results demonstrate that OGPAL achieves more accurate and comprehensive fault cause identification than both RAG and RGCN baselines. By combining ontology-guided conceptualization with process-aware graph learning, it enables consistent and transferable fault cause identification across heterogeneous manufacturing lines, providing a solid foundation for LLM-driven diagnostic assistance in smart manufacturing.

\section{Discussion}
\label{sec:discussion}

The case study demonstrates that OGPAL improves fault cause identification compared with retrieval-based and graph-learning baselines. However, the results should be interpreted considering several factors. Although the dataset contains 1,213 FMEA records, the diversity of process patterns is limited to 11 automated manufacturing contexts. Therefore, the model may not fully capture uncommon equipment configurations or highly line-specific causal pathways in different contexts, such as injection molding and human-assisted assembly. The quality of the target process model is also critical. If the process flow omits relevant operations, the process-aware scoring function may incorrectly prioritize causes that are semantically plausible but physically irrelevant. Similarly, regarding the expressiveness of the knowledge representation, the proposed ontology captures major concepts, but more detailed and diverse representations of physical interactions may be required for more precise diagnosis.

F1@20 of 0.524 indicates that while OGPAL significantly improves over baselines, there is still substantial room for improvement in fault cause retrieval. The errors in the fault cause retrieval results can be attributed to three main factors. First, some valid causes are underrepresented or absent in the training FMEA records, making them difficult to infer through graph learning. Second, some causes are described at different levels of abstraction across lines, such as generic ``position error'' versus process-specific ``camera misalignment'', which can lead to partial semantic matching but incorrect process grounding. Third, the current process representation does not always contain sufficient information about which components are active in each operation. For example, causes related to cameras may be ranked for a placement-related failure if visual inspection knowledge is strongly connected to misalignment concepts in the KG. These failure cases suggest that future work should incorporate richer process-component constraints and uncertainty-aware ranking rather than relying only on learned graph connectivity and process order.

From an industrial application perspective, OGPAL should be viewed as a diagnostic assistance tool rather than an autonomous decision-making system. The achieved F1@20 of 0.524 indicates that the framework can reduce the search space and present operators with a more relevant set of candidate causes than the baselines. However, this level of performance is not sufficient to replace expert judgement in quality-critical settings. Its practical value lies in supporting early-stage troubleshooting, knowledge reuse across lines, and systematic recall of possible causes that may otherwise be overlooked. In deployment, the output should therefore be integrated into a human-in-the-loop workflow, where operators validate, reject, or add candidate causes.

The framework also supports incremental knowledge updating when a new failure is manually solved. The newly observed failure description, confirmed cause, and related process step can be added to the KG as new nodes and relations for future retrieval and graph learning. Periodic retraining or fine-tuning of the RGCN can further propagate this new knowledge, enabling the KG to evolve with accumulated maintenance experience. However, when FMEA records are limited, the model may struggle to infer causes for novel failure scenarios that are not well represented in the training data. In such cases, the framework may still provide some relevant candidates based on shared ontology concepts and process flow, but the predictions are expected to be less reliable. This highlights the importance of continuously enriching the FMEA dataset and ensuring that it captures a wide range of fault phenomena for effective knowledge transfer across lines.

Finally, the proposed ontology-guided approach differs from traditional rule-based expert systems. Rule-based systems encode explicit if--then rules defined by experts, which makes them interpretable but difficult to scale across heterogeneous lines with different terminology and process configurations. In contrast, the proposed framework uses the ontology to constrain and generalize domain concepts, while graph learning captures statistical patterns from accumulated FMEA records. Thus, the ontology provides a structured semantic layer that enables LLM-driven extraction, KG integration, and data-driven inference rather than a fixed rule base. This makes the framework more flexible than purely rule-based systems, while still retaining domain structure needed for engineering interpretability.

\section{Conclusion}
\label{sec:conc}

This study presented a method to enhance the reusability of FMEA knowledge for fault cause identification across heterogeneous manufacturing lines. Through ontology- and LLM-guided conceptualization and process-aware graph learning, OGPAL captures both semantic relationships and process flow dependencies within a unified KG. In a case study on an automotive pressure sensor assembly line, the method demonstrated consistent improvements over a state-of-the-art RAG approach and an RGCN baseline in terms of fault cause retrieval accuracy, measured by precision, recall, F1 score, and nDCG at $n$. Ablation studies confirmed that manufacturing-domain conceptualization plays a central role in enabling knowledge transfer, while process-aware learning complements it by ensuring consistency with process flow. Additionally, the integration of heterogeneous FMEA data from multiple lines was shown to be crucial for capturing diverse fault phenomena and improving generalization. In addition to fault cause retrieval performance, we quantitatively evaluated the quality of ontology-guided LLM-driven entity extraction, which forms the foundation of the proposed knowledge graph. The results indicate that core operational, structural, and fault elements can be reliably extracted under ontology constraints. These findings demonstrate that the approach can provide more accurate and transferable candidate causes to support knowledge-based diagnostic reasoning in automated manufacturing systems. 

However, the study also has limitations primarily related to the scope of knowledge representation and evaluation. OGPAL relies on available descriptions of the target process flow, which may not always contain sufficient granularity to capture all relevant components and actions. Future work could explore richer process representations by incorporating additional sources such as sensor data or digital twin models. Additionally, while the LLM-based extraction method significantly reduces manual effort in knowledge acquisition, its outputs may still require limited human oversight to ensure semantic accuracy. Further refinement of prompt design and the incorporation of human-in-the-loop feedback mechanisms could enhance reliability. In addition, the current evaluation focused on a limited set of manufacturing lines and fault scenarios. Future studies should extend validation to a broader range of heterogeneous manufacturing contexts and failure modes, while incorporating RPNs to support risk-based prioritization of candidate causes. Future work should also explore the integration of agentic RAG, where iterative reasoning and tool-based retrieval could further enhance the adaptability of fault diagnosis, particularly in handling complex or previously unseen failure scenarios.

\section*{Acknowledgment}

This research was supported by JST SPRING (Grant Number JPMJSP2108) and by the New Energy and Industrial Technology Development Organization (NEDO) under project JPNP18002. Sho Okazaki acknowledges, in a personal capacity, the scholarship support of the SATOMI Scholarship Foundation. We also wish to thank Mr. Toshinori Yasui (DENSO CORPORATION) and Mr. Kenshu Sato (KS Consultant) for their expert advice on the experimental design within the manufacturing domain.

\appendix

\section{Entity Extraction Algorithm}
\label{sec:appendix_entity_extraction}
\begin{algorithm}
\caption{Entity extraction}
\label{alg:entity_extraction}
\begin{algorithmic}[1]
\Require sentences $\mathcal{S}$; ontologies $(O_{\mathrm{act}}, O_{\mathrm{st}}, O_{\mathrm{comp}}, O_{\mathrm{param}})$; shortlist size $k$
\ForAll{$s \in \mathcal{S}$}
  \State $(act, st, comp, param) \gets \mathrm{LLM\_Extraction}(s,$ \textsc{SystemPrompt}, \textsc{JsonSchema}$)$
%   \State \textbf{DeduplicateTokens}$(act, st, comp, param)$
  \ForAll{$x \in \{act, st, comp, param\}$}
    \If{$x \neq \varnothing$}
      \State $O_x \gets$ appropriate ontology/subtree
      \State $C \gets \mathrm{ShortlistByStringMatch}(x, O_x, k)$
      \State $y \gets \mathrm{LLM\_SelectID}(x, C, \texttt{allow\_new}{=}\texttt{True})$
      \State $id(x) \gets (y \neq \texttt{NEW}) \; ? \; y \; : \; (\texttt{NEW}, \mathrm{ParentFrom}(C), \mathrm{SuggestLabel}(x))$
    \Else
      \State $id(x) \gets \varnothing$
    \EndIf
  \EndFor
  \State \textbf{EmitRow}$(s, act, id(act), st, id(st), comp, id(comp), param, id(param))$
\EndFor
\end{algorithmic}
\end{algorithm}

\begin{lstlisting}[language=,caption={Prompt Excerpt for LLM\_Extraction}, label={lst:llm_extraction}]
System Prompt:
You are a slot extractor familiar with terminology used in manufacturing sites.
Given a short Japanese description, extract the primary terms for Action, State, and Object (= Component, Parameter) without normalization.
The output must always be in JSON format. Use null when not applicable. 
Do not assign the same word to multiple slots (no duplicate assignments).

Decision rules (important):
- Expressions ending with "function / performance / parameter / setting value / characteristic" are prioritized as Parameter.
- Equipment or device names (e.g., robot, sensor, conveyor, chuck, cylinder) are classified as Component.
- Terms such as "decrease / increase / excessive / insufficient / degradation / fracture / misalignment / slip / contamination / scratch / foreign substance adhesion / omission" are classified as State.
- Operational or processing terms (e.g., conveyance, gripping, discharge, inspection, cutting, bonding, etc.) are classified as Action.

User Prompt:
Description: {sentence}
Respond strictly following the JSON schema below:
{
 "action": {"text": string|null},
 "state": {"text": string|null},
 "object": {
   "component": {"text": string|null},
   "parameter": {"text": string|null}
 }
}
\end{lstlisting}

\begin{lstlisting}[language=,caption={Prompt Excerpt for LLM\_SelectID}, label={lst:llm_select_id}]
System Prompt:
You are a classifier for manufacturing process ontology.
You must always return the output in JSON format: {"element": "<ID or NEW>"}.
Select exactly one most appropriate ID from the candidates.
Do not include any extra text or explanations in the output.
If none of the candidates are appropriate, return "NEW".
Only when returning "NEW", use the format {"element":"NEW","new_parent":"<parent ID>","new_label":"<proposed label>"}.
Select exactly one closest parent from the candidates as new_parent.

User Prompt:
Input sentence:
\"\"\"{sentence}\"\"\"

Candidates (ID -> Label):
{id_to_label_block}

Example output:
{{"element": "A-010"}}  or  {{"element":"NEW","new_parent":"A-002","new_label":"Shooter conveyor"}}
\end{lstlisting}

 \bibliographystyle{elsarticle-num} 
 \bibliography{references_revision1}





% \end{thebibliography}
\end{document}